\documentclass[pra,aps,twocolumn,showpacs,superscriptaddress,floatfix]{revtex4}
\usepackage[ansinew]{inputenc}
\usepackage{amssymb}
\usepackage{graphicx}
\usepackage{amsmath}
\usepackage{mhequ}
\def\eq#1{\begin{equs}#1\end{equs}}
\def\entreg#1{``#1''}
\def\cbin#1#2{\binom{#1}{#2}}

\def\fig#1#2{%
\begin{figure}[htbp]
\begin{center}
\includegraphics[width=\columnwidth]{#1}
\caption{#2}
\label{#1}
\end{center}
\end{figure}
}
\def\dc{p^{\textrm{dc}}}
\def\pphot{p^\textrm{phot}}
\def\Pcount{p^\textrm{count}}
\def\gtwo{g^{(2)}(0)}
\def\Mn{M(n)}
\def\pn{p(n)}
\def\pA{p_A}
\def\pB{p_B}
\def\pzero{p(0)}
\def\pAB{p_{AB}}
\def\Mzero{M(0)}
\def\Pzero{P(0)}
\def\Pone{P(1)}
\def\Ptwo{P(2)}
\def\eff{\beta}
\def\effA{\eff_A}
\def\effB{\eff_B}
\def\dcA{\dc_A}
\def\dcB{\dc_B}

\def\effAeffB{\eff_{A/B}}
\def\dcAdcB{\dc_{A/B}}
\def\pAm{p^\textrm{m}_A}
\def\pBm{p^\textrm{m}_B}
\def\pABm{p^\textrm{m}_{AB}}

\def\net{^{\textrm{net}}}

\begin{document}
\title{High quality asynchronous heralded single photon source at telecom wavelength}
\author{Sylvain Fasel}
\email{sylvain.fasel@physics.unige.ch}
\affiliation{Group of Applied Physics, University of Geneva\\
20, rue de l'École de Médecine\\
CH-1211 Geneva 4, Switzerland}
\author{Olivier Alibart}
\affiliation{Laboratoire de Physique de la Matière Condensée\\
CNRS UMR 6622, Université de Nice--Sophia Antipolis\\
Parc Valrose, 06108 Nice Cedex2, France}
\author{Alexios Beveratos}
\author{Sébastien Tanzilli}
\author{Hugo Zbinden}
\affiliation{Group of Applied Physics, University of Geneva\\
20, rue de l'École de Médecine\\
CH-1211 Geneva 4, Switzerland}
\author{Pascal Baldi}
\affiliation{Laboratoire de Physique de la Matière Condensée\\
CNRS UMR 6622, Université de Nice--Sophia Antipolis\\
Parc Valrose, 06108 Nice Cedex2, France}
\author{Nicolas Gisin}
\affiliation{Group of Applied Physics, University of Geneva\\
20, rue de l'École de Médecine\\
CH-1211 Geneva 4, Switzerland}


\begin{abstract}
We report on the experimental realization and characterization of an asynchronous heralded single photon source based on spontaneous parametric down conversion. Photons at 1550\,nm are heralded as being inside a single-mode fiber with more than 60\% probability, and the multi-photon emission probability is reduced by a factor up to more than 500 compared to Poissonian light sources.
These figures of merit, together with the choice of telecom wavelength for the heralded photons are compatible with practical applications needing very efficient and robust single photon sources.
\end{abstract}
\maketitle
With the present development of quantum communication and computing technologies, including quantum key distribution \cite{cryptoreview} and quantum teleportation \cite{teleportbase} the interest for true single photon sources is rising. Many different implementations have been investigated, using single molecule or atom excitation \cite{molecul1,molecul2,molecul3,atom}, color centers in diamonds \cite{diamondwein,alexios2}, quantum dots \cite{moreau,yamamoto,shieldsscience,qdotEPFL1,qdotEPFL2} or pulsed parametric down conversion sources \cite{pittman,olivier}. All theses solutions have various advantages and tradeoffs between high purity and efficient single photon production, repetition rate, wavelength of the photons, and ease of use. The aim of this paper is to show that a spontaneous parametric down conversion (SPDC) source made of a bulk non-linear cristal at room temperature and a simple basic optical setup can be used to herald single photons at telecom wavelength in a very efficient way (see figure \ref{schemaHS}). The photons are also directly available in a standard single-mode telecom optical fiber, making this source a good choice for quantum communication applications such as scalable quantum networks. In this context, the term \entreg{heralded} means that photons are not generated on demand, but instead an electric signal announces the presence of a photon in a fiber. Indeed, as photons are created in pairs, the detection of one photon can be used to announce the presence of the complementary photon \cite{heralding}.

The source presented in this paper is an asynchronous heralded single photon source (AHSPS) because the heralding signals are not synchronized with a periodic clock, the SPDC pump being continuous. It exhibits a very good probability of producing one photon and a very low probability of producing more than one photon per heralding signal. The latter probability depends on the pump power applied to the crystal, and several measurements are presented to characterize this dependency.
\fig{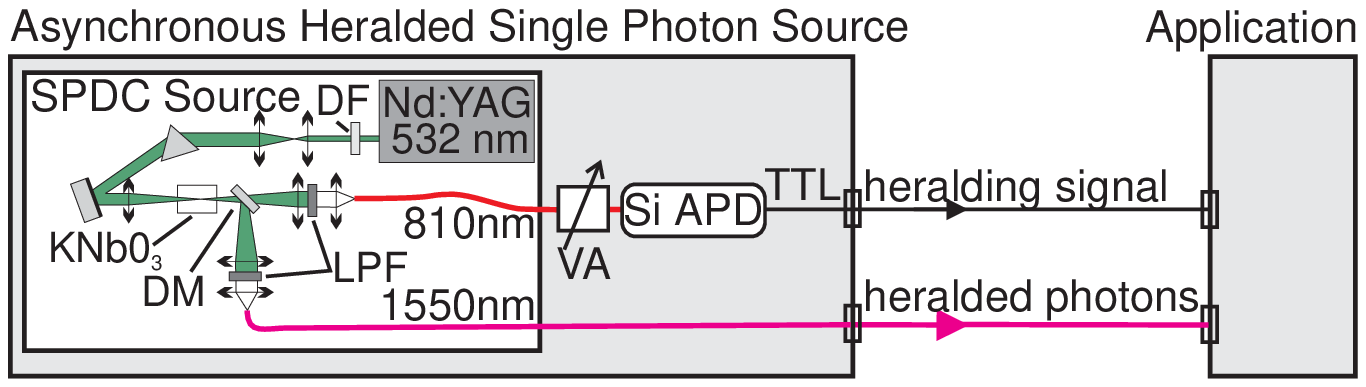}{Schematic of the asynchronous heralded single photon source (DM: dichroic mirror, DF: neutral density filter for pump attenuation, LPF: low-pass pump filters, VA: variable optical fiber attenuator)}

Our AHSPS is made of two main parts, as depicted in figure \ref{schemaHS}. The first consists in the SPDC photon pair creation stage which consists in a type I KNb0$_3$ bulk non-linear crystal, pumped with a 532\,nm continuous-wave laser (with maximum power of about 60mW). On one hand we require our source to produce single photons for long range telecom applications, and on the other hand we need the complementary photons to be passively and efficiently detected for heralding purpose. The phase matching condition is therefore chosen to produce 1550--810\,nm photons pairs, with the photons at 1550\,mn featuring a 6.9\,nm spectral width. A dichroic mirror is used to separate the 810\,nm photons from the 1550\,nm photons into two distinct beams, enabling one to couple these two wavelength using separated optimized optics into single-mode fibers. More details about the core of this source can be found in \cite{cryptoribordy,cryptofasel}. The 1550\,nm single mode fiber is the heralded single photons output port.

The second part of the AHSPS is made of a actively quenched silicon avalanche photodiode (APD) (EG\&G) connected to the 810\,nm single-mode fiber. The TTL detection signal from the silicon counter is the asynchronous heralding electric signal. For characterization, the optical variable attenuator in front of the silicon APD can be used to reduce the heralding signal rate without changing the heralded photons statistics. For the experiment, the pump power is varied by inserting neutral density filters in the path of the pump laser beam.  

The heralded photon output port is connected to a test bench consisting of a gated detection Hanbury-Brown \& Twiss type setup \cite{HBT}. In order to characterize the source, the relevant parameters $\Pone$, $\Ptwo$ and $\gtwo$ have indeed to be measured. $\Pone$ is the probability of having exactly 1 photon (and $\Ptwo$ the probability of having more than 1 photons) in the photon output port per heralding signal and during a time corresponding to the detection gate width. Note that the Hanbury-Brown \& Twiss setup does not enable to discriminate between 2 and more photons events. By convention, we thus have $\Pzero+\Pone+\Ptwo=1$, where $\Pzero$ is the probability of having no photon at all per heralding signal. Finally, $\gtwo$ is the standard second-order autocorrelation function at time 0 \cite{gdeux1}.

The setup is made of a 50/50 fiber optical beam splitter and two InGaAs APDs operated in gated mode (idQuantique). Both of these detectors are triggered with the heralding signal of the source through a TTL fan-out. The effective gate width is about 2.5\,ns and is chosen with respect to the jitter of Si APD, which is about 1\,ns. A narrower gate width would not allow to detect all the heralded photons and would thus bias down the measured value for $\Pone$ and $\Ptwo$. The measured dark count probability per gate is $\dcA=35.1 \pm0.7\times10^{-6}$ for detector A and $\dcB=7.4\pm0.3\times10^{-6}$ for detector B. The dark counts of the AHSPS silicon detector (about 100\,Hz) artificially increase the probability of having no photons in the output port, but this effect is measured to be negligible in the present case. The detection states of both detectors are acquired using a computer by means of a digital input pc card (PCI-6533 DIO-32HS, National Instrument) and a labview software  (see figure \ref{schemaTB})
\fig{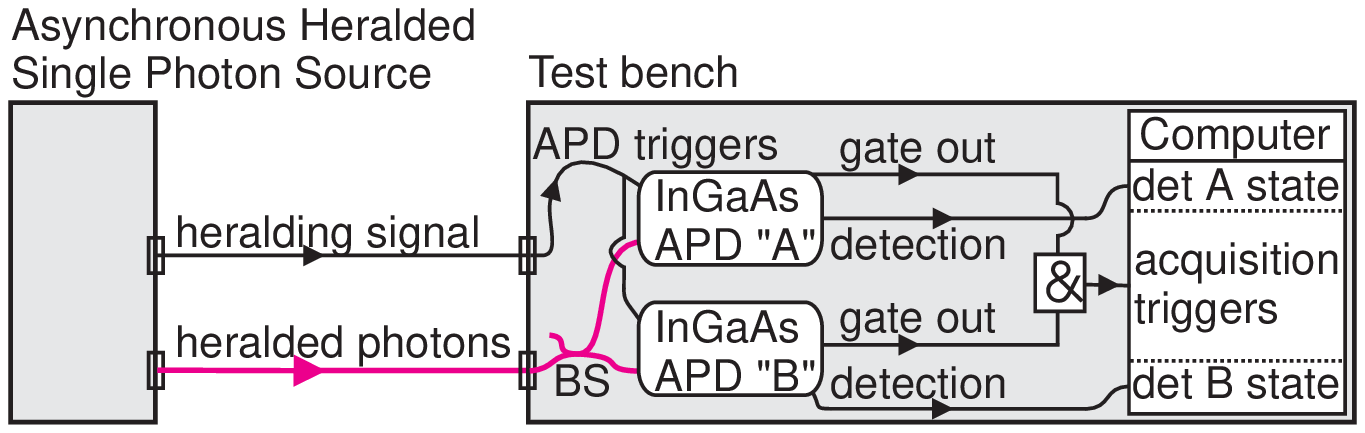}{Schematic diagram of the test bench. The beam splitter (BS) is considered as having a perfect 50/50 separation ratio and the fibers from it to the detectors are considered as lossless. This is possible because detectors efficiency are calibrated including losses in the beam splitter and in the fibers (see text).}
The triggering rate of the APDs is kept under 50\,kHz using the optical variable attenuator of the source. Note that this does not affect the quality of the source, but has to do with the test bench we use. Indeed, this low triggering rate and a 10$\mu$s external deadtime are used on both the InGaAs APDs to work in a regime where after-pulses counts are negligible.  Because of these deadtimes, the heralding signal of the source cannot be directly used to trigger the APDs state acquisition, but one must discard the cases where one or both detectors are disabled when a new trigger arrives. If not, an artificially lowered coincidence detection rate would be measured, and the quality of the source would be overestimated.  This is done by using a signal named "gate out" on figure \ref{schemaTB} which is synchronized with the effective application of bias voltage gate on a given APD, meaning that it is ready to detect heralded photons. These signals are combined with a logic AND gate to form the acquisition triggering signal. Upon reception of a trigger signal, the detection states (0 or 1) of the two APDs are acquired independently. The raw output is a binary file containing the detection state for each heralding signal effectively taken into account. 

Using attenuated laser pulses, the efficiencies of both detectors under these particular working conditions are calibrated including the losses in the beam splitter and in the fibers connecting the APDs. It is thus possible to consider that the beam splitter is a perfect 50/50 separator and that the fibers are lossless. The overall detection efficiencies are measured to be $\effA=8.4\pm0.1\%$ for detector A and $\effB=9.6\pm0.1\%$ for detector B. 

For the present setup, one can show that the $\gtwo$ parameter is equal to
\eq{
\gtwo\equiv\frac{\pAB}{\pA\,\pB}\cong\frac{2\Ptwo}{\Pone^2}\label{gtwo} 
}
where $\pAB$ is the experimental probability of having simultaneous detections in both detectors and $\pA$, $\pB$ the probabilities of having  a single detection in the given detector A or B. Note that, as expected, this quantity does not depend explicitly on the detection efficiency. Refer to the appendix for details on the rightmost part of the equality (\ref{gtwo}) and for details on obtaining $\Pone$ and $\Ptwo$ from the experimental data. $\pA$ and $\pB$ are obtained from the raw binary files by counting the number of registered detection for the given detector ($N_A$, $N_B$) divided by the total number of acquisition triggers ($N_t$). $\pAB$ is obtained by counting how many times both detectors have registered a simultaneous detection ($N_{AB}$) divided by the total number of acquisition triggers.
\begin{table*}
\caption{Experimental AHSPS parameters obtained with a detection gate width of 2.5\,ns}
\label{results}
\begin{ruledtabular}
\begin{tabular}{cccccc|ccc}
$P_{p}$&
$R_H$&
$N_t$&
$T$&
$N_A$&
\phantom{a}\ \,$N_{AB}$\ \,\phantom{a}&
$\Pone$&
$\Ptwo$&
$\gtwo$\\
\hline
49\,mW    & 845\,kHz&$ 6.0\times10^7  $&20\,min. &$16\times10^5$ &$1062$&$0.59$&$41\pm1\times10^{-4}$     &$22.6\pm0.5\times10^{-3}$\\
25\,mW    & 403\,kHz&$ 2.8\times10^7  $&20\,min. &$7.7\times10^5$ &$255$&$0.61$&$19\pm1\times10^{-4}$     &$10.2\pm0.5\times10^{-3}$\\
13\,mW    & 221\,kHz&$ 1.6\times10^7  $&20\,min. &$4.1\times10^5$ &$82$&$0.61$&$10\pm1\times10^{-4}$     &$5.3\pm0.5\times10^{-3}$\\
3.3\,mW    & 120\,kHz&$ 1.7\times10^7 $&12\,min. &$4.6\times10^5$ &$78$&$0.60$&$8.0\pm0.9\times10^{-4}$  &$4.4\pm0.5\times10^{-3}$\\
1.6\,mW    & 39\,kHz&$1.6\times10^7  $&40\,min. &$4.2\times10^5$ &$35$&$0.61$&$2.5\pm0.6\times10^{-4} $  &$1.4\pm0.3\times10^{-3}$\\
0.9\,mW   & 18\,kHz& $1.8\times10^7   $&85\,min. &$4.4\times10^5$ &$46$&$0.58$&$3.7\pm0.6\times10^{-4}$ &$2.2\pm0.4\times10^{-3}$\\
\end{tabular}
\end{ruledtabular}
\end{table*}

For the measurement the SPDC source is first aligned without neutral density pump filters, by optimizing the collection of the 810\,nm photons.  A second alignment is done by maximizing the ratio of detection at 1550\,nm over detection at 810\,nm through the adjustment of both 1550\,nm and 810\,nm photons collection. The highest ratio is usually found when the 810\,nm detection rate is about 90\% of the maximal rate. Different neutral density pump filters are introduced one after the other in the path of the pump beam. For each filter, the pump power is measured just before the crystal, and the heralding signal rate $R_H$ is recorded. When necessary, the variable attenuator is set to a non-zero value to keep this rate under 50\,kHz for the acquisition. The results are summarized in table \ref{results}. In this table, $P_{p}$ is the pump power and $T$ is the experimental duration of the acquisition. $\Pone$, $\Ptwo$  and $\gtwo$ are net (dark count substracted values) calculated from formulas developed in the appendix. These values are obtained from the averages of the slightly different results calculated using $N_A$ and $N_B$. As simultaneous detections are quite rare events, $N_{AB}$ is small. The statistical error on this quantity is the main source of error for $\Ptwo$ and $\gtwo$. It can be relatively high but it is kept under an acceptable level by acquiring for relatively long times. The error on $\Pone$ is negligible, because $N_A$ (and $N_B$) is large. The slight changes for $\Pone$ values (less than 2\% around the average value of $0.6$) are explained by the fact that the neutral density pump filters slightly deviate the pump beam. This causes a change in the phase matching condition and consequently a decrease of the coupling ratio from its nominal value. As $\Pone\gg\Ptwo$, $\Pone$ is to a very good approximation the heralded photons coupling ratio, which is sometimes called the heralding efficiency $H$ \cite{pittman}.
\fig{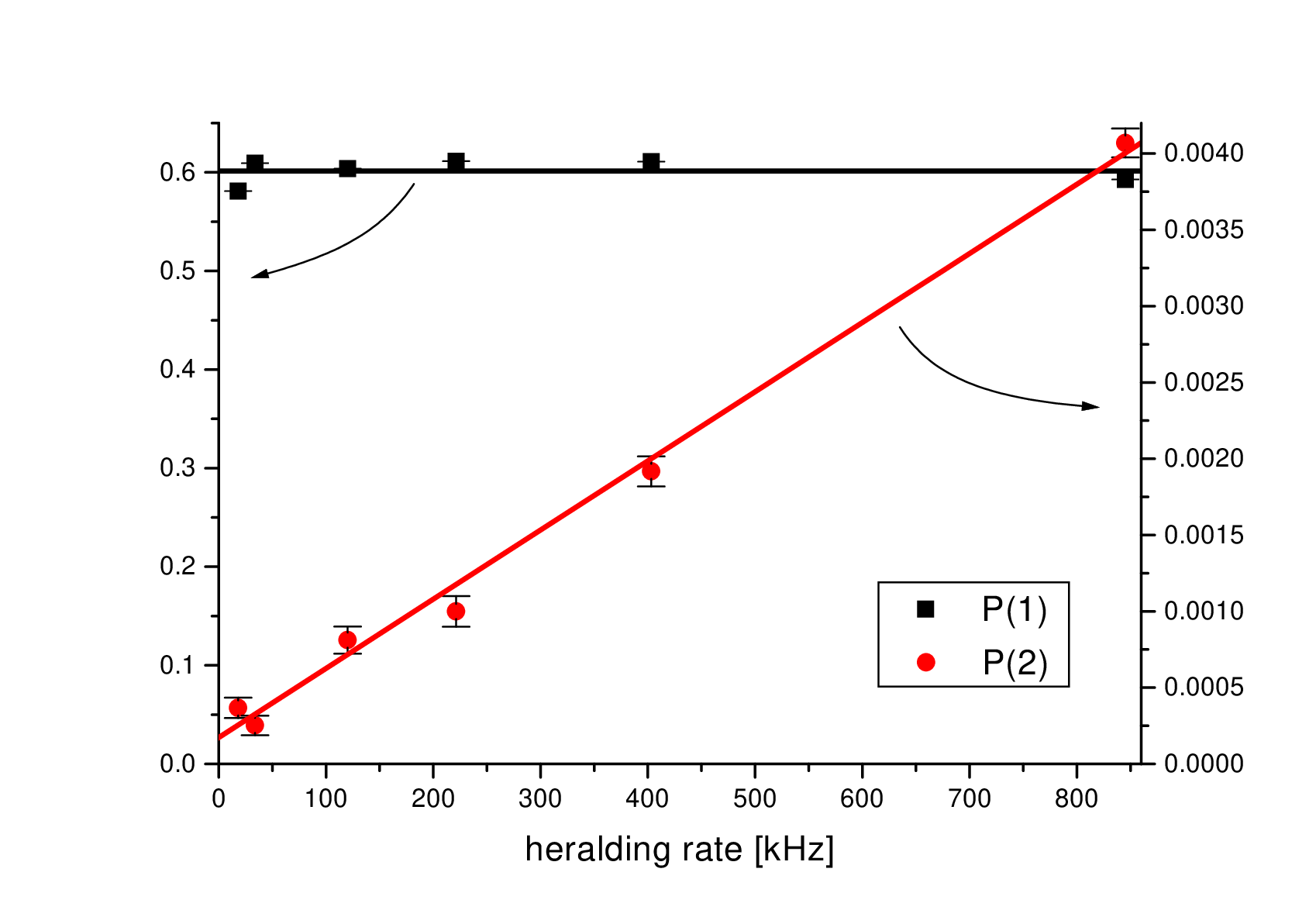}{$\Pone$ (left scale) and $\Ptwo$ (right scale) displayed as function of the heralding rate. The solid lines are: average for $\Pone$ and linear fit for $\Ptwo$. As expected, $\Ptwo$ is linearly dependent on heralding rate (and hence on the pump power), while $\Pone$ remains constant.}

These results should be compared to a Poissonian light source featuring the same value for $\Pone$. The main observation is that the multi-photon emission probability $\Ptwo$ of our source is reduced by a factor $1/\gtwo$ ranging from about 50 to more than 500, depending on the pump power. To the best of our knowledge, these are the highest factors reported so far.  In fact, $\gtwo$ and $\Ptwo$ can be arbitrarily low for low pump powers, i.e. heralding rates, as $\Ptwo$ is linearly dependent on the heralding rate (see figure \ref{graphP1P2singles}). Moreover, $\Ptwo$ can be reduced by lowering $\Pone$, i.e. misaligning the heralded photons coupling.  A better figure of merit is therefore the ratio of the heralded photons rate over $\gtwo$:
\eq{
F=\frac{\Pone R_H}{\gtwo}=\frac{\Pone^3 R_H}{2\Ptwo}}

$R_H$ and $\Ptwo$ are related to the physical parameters of the source as follows: $R_H$ is the number of photon pairs created by second multiplied by the probability to detect the heralding photon. $\Ptwo$ is the probability to find two photon pairs in a time interval $\Delta\tau$ knowing that a trigger photon was detected, which is half of the probability that a photon pair is created during $\Delta\tau$, multiplied by the probability that two photons are coupled into the output fibre, $\Pone^2$. We find
\eq{
R_H=q_H\,\eta_H\,\eta_{S}\,P_{p}\quad,\quad
\Ptwo=0.5\,\Pone^2\,\Delta\tau\,\eta_{S}\,P_{p}
} where $q_H$ is the coupling efficiency of the heralding photons, $\eta_H$ is the quantum efficiency of the heralding detector, $\eta_{S}$ the pair creation efficiency of the SPDC crystal and $\Delta\tau$ equivalent to the detection gate width of the heralded photons detection. We can now express $F$ by
\eq{
F=\frac{\Pone\, q_H\,\eta_H}{\Delta\tau}
}
which allows us to see which parameters should possibly be improved. We note that $F$ does not depend on the efficiency of the SPDC crystal. Thus, using more efficient pair generation, e.g. by using waveguides in periodically poled crystals \cite{olivier}, would not impact on the quality of this single photons source. However, the source can be improved by: i) Increasing $\Pone$ by using  better optics, coatings or alignments \cite{pittman}, or using a different photon pair creation technique as the Kerr effect inside optical fiber \cite{kumar}; ii) Improving the coupling and detection efficiency of the heralding detector; iii) Reducing the timing jitter of the heralding detector which would allow to use much shorter detection gate width $\Delta\tau$ for the heralded photons.

We found interesting to compare in a standard visual way the antibunching properties of the non-classical light coming from our source to other published results.  Therefore, we devise a way to use the standard technique used in the case of synchronous single photon source, even though our source is asynchronous. For synchronous single photon sources, the standard way is to use detection signals coming from the Hanbury-Brown \& Twiss setup as start and stop for a time to digital converter (TDC) device. Indeed, the time between successive possible creations and detections of photons is constant (fixed by a given clock signal), and such setups enable one to directly obtain histograms where bin $n$ represent the probability of having two detections in different detectors, that happened at an interval of $n$ triggers. For true single photon source, the central bin ($n=0$) is of course zero while the other bins behave as for a poissanian source. The deepness of this central dip allow one to have a intuitive information about the quality of the single photon source considered.

For the present source, single photons are created at random times because of the continuous pumping and spontaneous non-linear process used. The elapsed time between a detection in one detector and the other is therefore completely random, and for this reason it is not possible to use a TDC. Instead, the number of triggers between detection in different detectors is computed directly from the raw binary data by a home made software. Our reduction algorithm enables us to present standard histograms.

More precisely, the algorithm is the following: the program sequentially reads the raw data and registers in bins how many times a start and a stop were separated by $n$ triggers (this number is called $\Mn$ in the following). As a convention, $n$ is positive if the start was given by detector $A$ and the stop by detector $B$, and negative in the opposite case. The bin $n=0$ corresponds to the case for which two photons were detected simultaneously (i.e. for the same trigger). As for standard TDC, there could be some invalid start (another start for the same detector before a stop on the other) that will be ignored. Moreover the search for a stop is done only within a limited range: if no stop is found before a certain number of triggers (we used 100) the search is cancelled and a new start has to be found in the subsequent data. All the processes leading to the histogram construction are summarized on figure \ref{algo1}. Note that this technique could be possibly applied to synchronous sources too.
\begin{figure}[h]
\begin{center}
\begin{tabular}{l|c|c|l}
\cline{2-3}
            &det $A$&det $B$&\\
\cline{2-3}
            &\dots&\dots&\\
\cline{2-3}
no detection\quad & 0&0&\quad$n=-1$\\
\cline{2-3}
start at A\quad& 1&0&\quad$n=\phantom{+}0$\\
\cline{2-3}
no detection\quad         &0&0&\quad$n=+1$\\
\cline{2-3}
invalid start\quad&1&0&\quad$n=+2$\\
\cline{2-3}
no detection \quad           &0&0&\quad$n=+3$\\
\cline{2-3}
stop at B\quad& 0&1&\quad$n=+4$\\
\cline{2-3}            
        &\dots&\dots&\\
\cline{2-3}
\end{tabular}
\caption{Samples of raw data. Here, as an example, data reduction leads to a start and a stop separated by 4 triggers.}
\label{algo1}
\end{center}
\end{figure}

The theoretical form for $\Mn$, when $n\neq0$ is given by
\eq{
\Mn=C\,\pA\,(1-\pB)^{|n|}\,\pB\label{Mn}
} for $n>0$ and with $\pA$ and $\pB$ swapped for $n<0$, where $C$ is a normalisation constant which depends on the total number of triggers, on the maximal $n$ allowed, and on $\pA$ and $\pB$. Detections that do not occur simultaneously are indeed coming from different photons pairs and thus lead to independent detection events at detectors $A$ and $B$. The probability $\pn$ of having $n$ triggers between a start and a stop is thus given by $\pn=\frac{\Mn}{C}$. $C$ is obtained by fitting the experimental results with the formula above for $n\neq0$. Figure \ref{resultsraw} shows a typical result for the whole histogram created by the algorithm described above, using 100 as range limit value. The central bin (corresponding to the measured value for $\Mzero$) is nearly 0 indicating a very low $\pzero=\frac{\Mzero}{C}\cong\pAB$. The data are in a very good agreement with formula (\ref{Mn}) as shown by the fitted curve. The decreasing slope on both side of the center is related to the detection probabilities $\pA$ and $\pB$. Theses values are quite high in our experiment compared to others. Indeed, recall that $\pA$ and $\pB$ describe not only the detection efficiencies but also the collection efficiencies which are rather high in our case. Consequently, the slope in figure \ref{resultsraw} is quite steep and may look unfamiliar. 
\fig{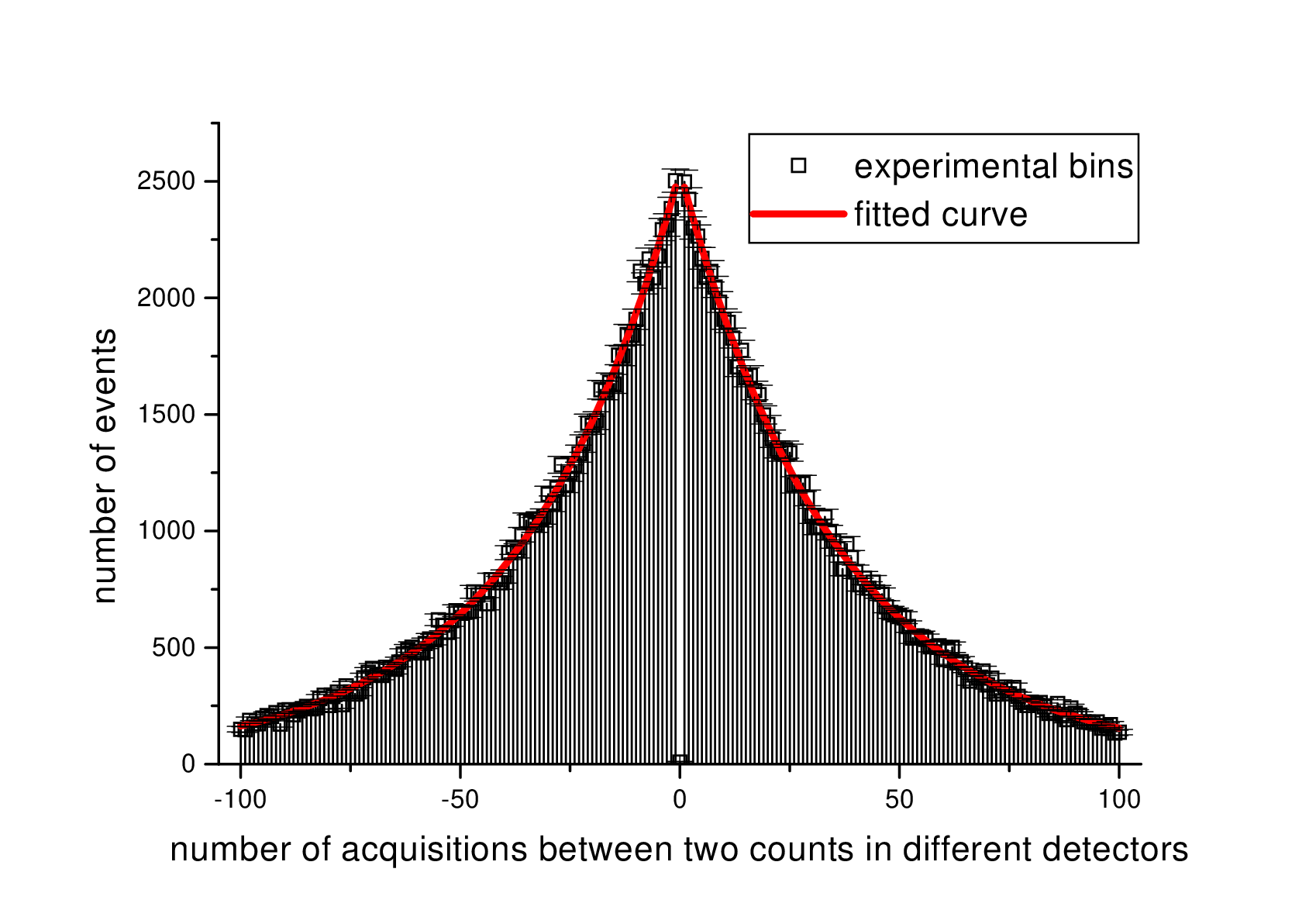}{Typical resulting raw histogram displayed as points over the full range allowed by the algorithm. The plain curve is obtained by fitting the experimental values with the formula (\ref{Mn}) and varying $C$.}

Another mean of presenting these data is to restrict the display to a limited bin range around the center, and to normalize the values in such a way that $\gtwo$ is directly readable on the graph. This is done by dividing all experimental values by $C\,\pA\,\pB$. The height of the central bin corresponds then directly to $\gtwo$ without dark counts substraction (see appendix for details on dark count substraction). Figure \ref{resultszoom} shows this representation for the same data as on figure \ref{resultsraw}.
\fig{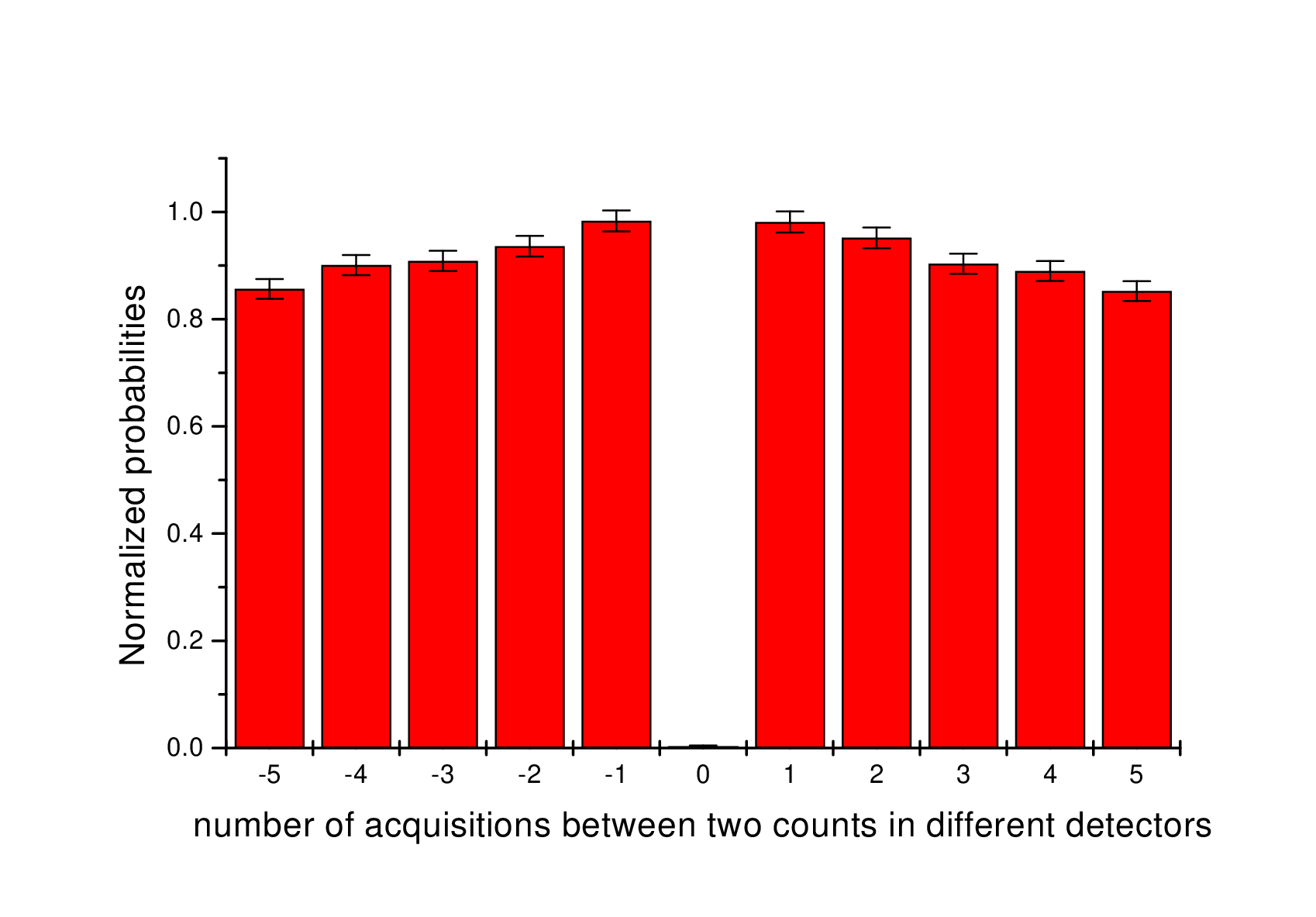}{Typical resulting normalized histogram, restricted to small values around the center. It is possible to infer the $\gtwo$ values from the central bin. This histogram correspond to acquisition made for 1.6\,mW of pump. Note that the bins are made using raw values (dark counts are not substracted) and that in this case $\gtwo=3\times10^{-3}$.}

Note that it is in principle possible to turn our source into a synchronous one. A first step would be to use a pulsed laser to pump the non-linear crystal. A second step, following \cite{fransonSHSPS}, would consists in a switched storage delay line (see figure \ref{AHSPStoSHSPS}). The latter would allow one to increase the probability of getting precisely one photon provided that the switch has a low enough insertion loss. The heralding signal is activated each time the storage loop is released (e.g. once every 10 pump pulse cycles), provided the Si-APD did register one photon.
\fig{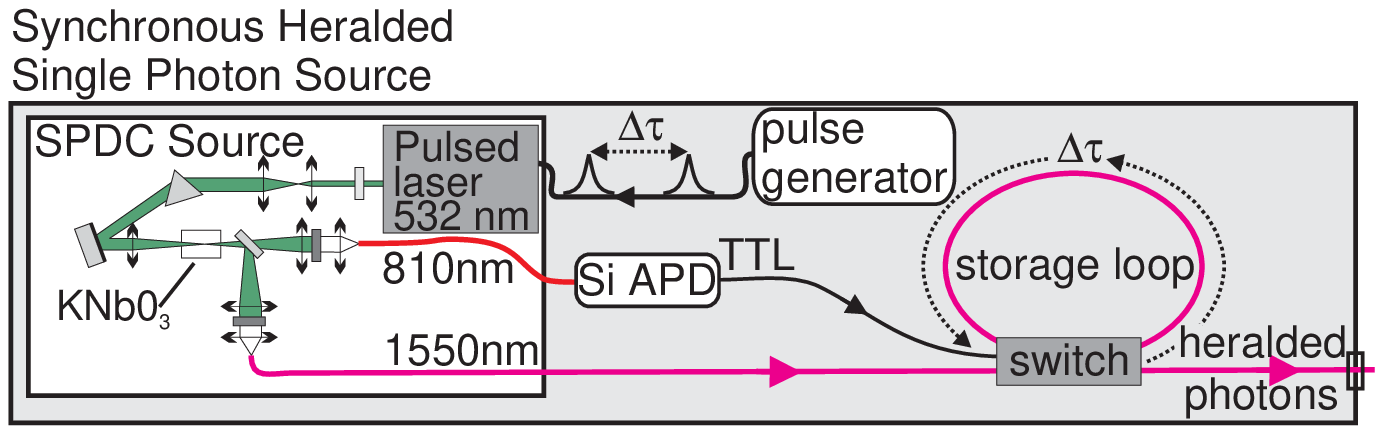}{Schematic for using a SPDC source for high efficiency synchronous single photons production}

We demonstrated that a simple and well known SPDC setup \cite{cryptoribordy,cryptofasel} can be used to implement a very efficient high quality single photon source. The properties of our source for different pump powers have been investigated, showing that good tradeoffs between very low $\gtwo$ and high enough repetition rate can be found, which are competitive with much more complex sources. Moreover, SPDC sources can be designed to produces photons at various wavelength to match particular needs. Our source produces telecom wavelength single photons that are already in a standard single mode telecom fiber, is simple to operate and works at room temperature without stabilization. It is therefore a very good candidate for practical real-world telecom applications for which single photons are required.

\appendix*
\section{}
Let us elaborate on the way we obtain the relevant source parameters after substraction of the dark counts for our setup (see figure \ref{schemaP}).
\fig{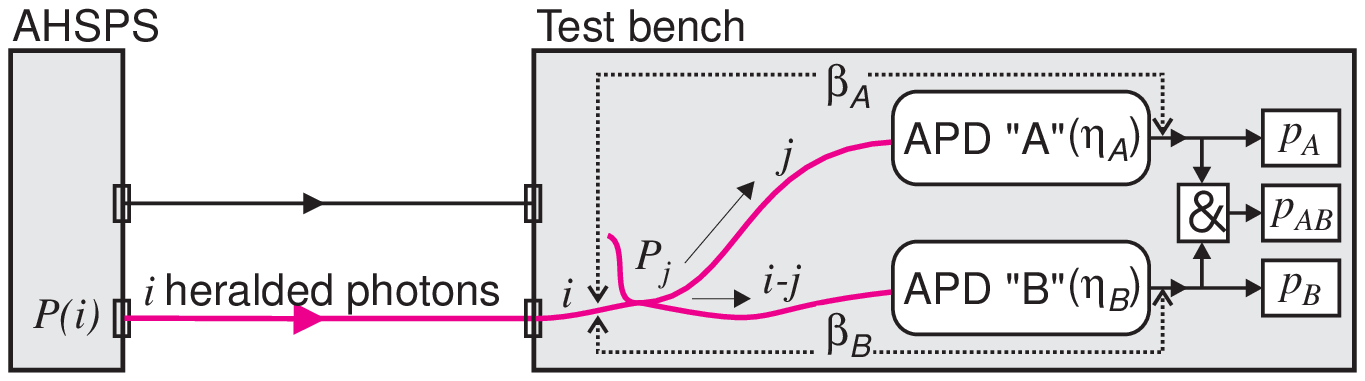}{Schematic of the quantities used for calculation of the source parameters.}

The probability $\Pcount$ for a single detector to register a count is given by
\eq{
\Pcount&=(1-\dc)\,\pphot+\dc\,(1-\pphot)\\
&=\pphot\,(1-2\dc)+\dc
}
where $\dc$ is the probability for the detector to register a dark count in the absence of photons, and with $\pphot=\sum_{j\geq0}\,P_j\,(1-(1-\eta)^j)$ the probability of detecting photons. Here $P_j$ ($\sum P_j=1$) represents the photon numbers distribution directly at the input of the detector which features a quantum efficiency $\eta$. At the input of the test bench, the beam splitter randomly separate $i$ incoming photons in the two arms. $P_j$ is thus the binomial probability that, having $i$ photons before the perfect beam splitter, $j$ photons reach a given detector ($j\leq i$) i.e. $P_j=\cbin{i}{j}\frac{1}{2^j}\frac{1}{2^{i-j}}=2^{-i}\cbin{i}{j}$. Using this $P_j$ into the above formula for $\pphot$, and using $\sum_j^i\cbin{i}{j}x^j=(1+x)^i$, $\pphot$ reduces to $\pphot=1-(1-\frac{\eta}{2})^i$.
Therefore, whenever a heralding signal announces the presence of a multi-photon state (with probability of having $i$ photons given by $P(i)$, $\sum\,P(i)=1$) at then input of the beam splitter, $\pA$ (the expression for $\pB$ is similar) is expressed by
\begin{equation}
\pA=\sum_{i=0}^{\infty}P(i)\left\{\left(1-2\dcA\right)\left(1-\left(1-\frac{\effA}{2}\right)^i\right)+\dcA\right\}\quad\label{pA_B}
\end{equation}
Here $\eta$ is substituted by $\effA$, the overall efficiency detection including fiber losses and quantum efficiency. 

In the same manner, the joint probability $\pAB$ can be calculated to be
\minilab{pAB}
\eq{
\pAB=&\sum_{i=0}^{\infty}P(i)\sum_{j=0}^i\cbin{i}{j}\frac{1}{2^i}
\Big{\{}\left[(1-(1-\effA)^j)\,(1-2\dcA)+\dcA\right]\\
&\times\left[(1-(1-\effB)^{(i-j)})\,(1-2\dcB)+\dcB\right]\Big{\}}\\
=&\sum_{i=0}^{\infty}P(i)\Bigg{\{}\left(1-\left(1-\frac{\effA}{2}\right)^i-\left(1-\frac{\effB}{2}\right)^i\right.\label{pAB:a}\\
&\left.+\left(1-\frac{\effA+\effB}{2}\right)^i\right)\,(1-2\dcA)\,(1-2\dcB)\tag{\ref{pAB:a}}\label{pAB:atag}\\
&+\left(1-\left(1-\frac{\effA}{2}\right)^i\right)\,(1-2\dcA)\,\dcB\label{pAB:b}\\
&+\left(1-\left(1-\frac{\effB}{2}\right)^i\right)\,(1-2\dcB)\,\dcA\tag{\ref{pAB:b}}\label{pAB:btag}\\
&+\dcA\,\dcB\Bigg{\}}\label{pAB:d}
}
The lines (\ref{pAB:a}) correspond to the cases for which at least one photon is detected simultaneously in each detectors, and are consequently non-zero only for $i>1$. The lines (\ref{pAB:b}) are the terms corresponding to the cases for which at least one photon is detected by one detector while the other one registers a dark count. These expression are therefore non-zero only for $i>0$. The last line (\ref{pAB:d}) corresponds to the case for which both detectors register dark counts, and is of course independent of $i$.

Assuming $\Pzero+\Pone+\Ptwo=1$ in (\ref{pA_B}) and (\ref{pAB}) (which is similar to neglecting terms for $i>2$), and solving for $\Pone$ and $\Ptwo$ ($\Pzero=1-\Pone-\Ptwo$), it is possible to get values for these quantities from a function of $\pAm$, $\pABm$, $\effAeffB$ and $\dcAdcB$ ($^\textrm{m}$ distinguish measured values for a parameters). $\Pone$ and $\Ptwo$ were also calculated from $\pBm$ and $\pABm$ and the final published values are the averages. These values take directly into account the efficiencies and the dark counts and are thus as close as possible to the AHSPS parameters.

The net values for $\pA$, $\pB$ and $\pAB$, denoted with $\net$, are obtained from
\eq{
\pA\net=\pA|_{\dcA=0}\quad&,\quad
\pAB\net=\pAB|_{\dcA=\dcB=0}
}
by evaluating the right members of the equalities using the value for $\Pone$ and $\Ptwo$ calculated from $\pAm$, $\pABm$, $\effAeffB$ and $\dcAdcB$ (or $\pBm$ instead of $\pAm$ for $\pA\net$). These values are used to calculate the net $\gtwo$ of table \ref{results} using \ref{gtwo}.

A simplified case with $P_2\ll P_1$ and with $\dcA=\dcB=0$ (no noise) and $\effA=\effB=\eff$ (these conditions correspond to a good approximation to our experimental results), leads to:
\eq{
\pAm=\pBm=\Pone\,\frac{\eff}{2}\quad&,\quad
\pABm=\Ptwo\,\frac{\eff^2}{2}\\
\intertext{and thus}
\Pone=2\,\frac{\pAm}{\eff}=\frac{\pAm+\pBm}{\eff}\quad&,\quad
\Ptwo=2\,\frac{\pABm}{\eff^2}
}
In this case we have
\eq{
\gtwo&=\frac{\pABm}{\pAm\,\pBm}=\frac{\pABm}{(\pAm)^2}=2\frac{\Ptwo}{\Pone^2}
}
and thus we obtain the well known expression for $\gtwo$ which enable one to compare the present source with a Poissonian source (as $\gtwo=1$ for such light sources).

\begin{acknowledgments}
Financial support by the Swiss NCCR Quantum Photonics is acknowledged.
\end{acknowledgments}

\end{document}